\documentstyle[12pt]{article}

\begin{document}
\begin{titlepage}
\today          \hfill
\begin{center}
\hfill    LBNL-42271 \\
\hfill    OITS-659 \\

\vskip .05in

{\large \bf  GUT and SUSY Breaking by the Same Field}
\footnote{This work was supported 
in part by the Director, Office of Energy
Research, Office of High Energy and Nuclear Physics, Division of High
Energy Physics of the U.S. Department of Energy under Contract
DE-AC03-76SF00098 and in part by DOE grant number DE-FG03-96ER40969.}
\vskip .15in
Kaustubh Agashe \footnote{email: agashe@oregon.uoregon.edu} 
\vskip .1in
{\em
    Theoretical Physics Group\\
    Lawrence Berkeley National Laboratory\\
    University of California,
    Berkeley, California 94720 \\}
and \\
{\em
Institute of Theoretical Science \\
5203 University
of Oregon, Eugene, Oregon 97403-5203 \footnote{address after 
September 1, 1998.}\\}
\end{center}

\vskip .05in
            
\begin{abstract}
We present a model in which the same modulus field breaks both SUSY 
and a simple
GUT
gauge group down to the SM gauge group. 
The modulus is stabilized by the inverted hierarchy mechanism in a 
perturbative
region so that the model is calculable. 
This is the first example of this kind in the 
literature. All mass scales (other than the Planck
scale) are generated dynamically.       
In one of the models doublet-triplet splitting is achieved naturally
by the sliding singlet mechanism while another model requires fine tuning.   
The gauge mediation contribution to the right handed slepton (mass)$^2$
is negative. But, for the modulus vacuum expectation value close to the
GUT scale, the supergravity contribution to the slepton (mass)$^2$
is comparable to the gauge mediation contribution and thus a realistic
spectrum can be attained. 
\end{abstract} 


\end{titlepage}







\newpage
\renewcommand{\thepage}{\arabic{page}}
\setcounter{page}{1}

\section{Introduction}
One of the central issues in studying supersymmetric
extensions of the Standard Model (SM) is how to break 
supersymmetry (SUSY) and mediate SUSY breaking
to the 
{\em s}particles. In models of dynamical SUSY breaking, 
SUSY is broken by the non-perturbative effects of a gauge group. 
Thus, the SUSY breaking scale is
related to the energy scale at which some gauge group becomes
strong and, in turn, to the Planck scale by
dimensional transmutation. For the mediation of SUSY breaking to the 
sparticles, two mechanisms have been discussed in the literature
-- gravity and SM gauge interactions.

The measurements of $\sin ^2 \theta _W$ are in very good agreement with
the predictions of SUSY grand unified theories (GUT's).
This has led to a lot of interest in SUSY GUT's.
One of the 
important issues in
SUSY GUT's is the origin of the energy scale $\sim 2 \times 10^{16}$ GeV
at which the GUT symmetry breaks
down to the SM.

There have been efforts to generate the GUT scale dynamically.
In the models of Cheng \cite{cheng} and Graesser \cite{graesser}, the
GUT scale is related to the dynamical scale of a gauge group, but 
SUSY breaking is unrelated to GUT symmetry breaking,
{\it i.e.,} there is a separate dynamical scale for SUSY breaking and
GUT symmetry breaking.

In the models of Goldberg \cite{goldberg}, Kolda and Polonsky \cite{kolda}
and Chacko, Luty and Ponton \cite{luty}, there is a connection between
GUT and SUSY breaking.
However, there are two different sectors (and potentials) for GUT and
SUSY breaking, but with related parameters (and one dynamical scale). Once SUSY is broken in one sector, a potential is
generated for a field 
in another sector determining the GUT scale.
In other words, in these models, the field breaking the GUT
symmetry/determining the GUT scale is different from the field breaking 
SUSY. 

In the model of Hirayama, Ishimura and Maekawa \cite{jap}, 
the field breaking SUSY and GUT
symmetry is the same. However,
the GUT gauge group is
$SU(5) \times SU(3) \times SU(2) \times U(1)$, which is 
{\em not} a simple gauge group.\footnote{Thus, the unification of the 
SM gauge couplings
at $\sim 2 \times 10^{16}$ GeV is not an automatic consequence of the 
model.} 
Also an assumption about a non-calculable
K\"ahler potential is required for the model to work. 

In this paper, we present a model in which not only are SUSY breaking and 
GUT symmetry breaking related, but the {\em same} field breaks
both SUSY and a GUT gauge group down to the SM gauge group.
However, unlike the model of reference \cite{jap}, the GUT gauge group is
{\em simple}. 
The well known inverted hierarchy mechanism is used to generate
a local minimum for the modulus field in a perturbative region,
thus making the model calculable, unlike the model of 
reference \cite{jap}. 
There are no dimensionful parameters in the model other than the 
Planck scale.
The mediation of SUSY breaking to the sparticles is by a combination of 
gravity and SM gauge interactions.

\section{General Structure}
The gauge group of the model is:\footnote{This model was used
in \cite{dim1} as a model of gauge mediation. However, in \cite{dim1}, 
the SM was an 
additional gauge group, {\it i.e.}, it was 
not embedded in the $SU(6)$ gauge symmetry.} 
\begin{equation}
SU(6)_{GUT} \times SU(6)_S
\end{equation}
and the particle content is
\begin{eqnarray} 
\Sigma & \sim \;& ({\bf 35},{\bf 1}) \nonumber \\
Q & \sim
\; &({\bf 6},{\bf 6}) \nonumber \\
\bar{Q}
& \sim
\; & ({\bf \bar{6}},{\bf \bar{6}}). 
\end{eqnarray}
The superpotential is
\begin{equation} 
W_1 = \lambda_Q \Sigma Q \bar{Q} + \frac{\lambda_{\Sigma}}{3} \Sigma ^3.
\label{w1}
\end{equation}
$\Sigma ^3$ lifts all flat directions in $\Sigma$ except 
tr $\Sigma ^2$ \cite{dim1}
along which the vacuum expectation value (vev) of $\Sigma$, upto
$SU(6)_{GUT}$ rotations, is\footnote{We use the 
normalization tr $T_a T_b = 1/2 \; \delta _{ab}$, where the $T$'s are the 
generators for the fundamental representation of a gauge group.} 
\begin{equation}
\langle \Sigma \rangle = \frac{v}{\sqrt{12}} \; 
\hbox{diag}[1,1,1,-1,-1,-1].
\label{sigmavev}
\end{equation} 
This can be seen 
as follows.
The vev of $\Sigma$ breaks $SU(6)_{GUT}$ to
$SU(3) \times SU(3) \times U(1)$. 
The resulting Nambu-Goldstone fields, with their $SU(3) \times SU(3)$ 
quantum numbers, are:
\begin{equation}
{\bf (3,\bar{3}) + (\bar{3},3)}.
\label{6to33}
\end{equation}
$\Sigma$ decomposes as:
\begin{equation} 
{\bf (3,\bar{3}) + (\bar{3},3) + (8,1) + (1,8) + (1,1)}.
\label{sigma33}
\end{equation} 
Thus, the ${\bf (3,\bar{3}) + (\bar{3},3)}$ components of $\Sigma$ 
are eaten 
in the gauge symmetry breaking. The ${\bf (1,8) + (8,1)}$ components
get a mass from the 
$\Sigma ^3$ term and the ${\bf (1,1)}$ component is the flat direction.
Thus, far out along this flat direction, $Q,\bar{Q}$ and all components of
$\Sigma$ other than the flat direction are heavy.
The only light fields are the $SU(6)_S$ gauge field and the flat direction
parametrized by tr $\Sigma ^2$. We will denote the flat direction (both
the chiral superfield and the vev of it's scalar component) by $v$.
The dynamical scale, $\Lambda _L$, 
of the pure $SU(6)_S$
gauge theory is related to the dynamical scale, $\Lambda$, of 
the high energy $SU(6)_S$ by 
the matching relation at the mass of $Q,\bar{Q}$ 
(we assume $v \gg \Lambda$):
\begin{equation}
\left( \frac{\Lambda _L}{\lambda_Q v / \sqrt{12} } \right) ^{18} = 
\left( \frac{\Lambda}{\lambda _Q v / \sqrt{12}} \right) ^{12}.
\end{equation}
Gaugino condensation in the low energy $SU(6)_S$ generates the
superpotential:
\begin{equation}
W = 6 \Lambda_L^3 = \sqrt{3} \lambda_Q \Lambda ^2 v.
\end{equation}
Below the scale $\Lambda_L$, we have only the field $v$ with the above 
superpotential with $F_v =  \sqrt{3} \lambda_Q
\Lambda^2$. Thus, SUSY is broken and with 
a canonical K\"ahler potential, $v^{\dagger}v$, the vacuum energy is 
$3 \lambda _Q^2 
\Lambda^4$. The vev $v$ is undetermined at this level. To determine $v$, 
we need to include the corrections to the K\"ahler
potential of $v$. The dominant corrections, for $v \gg \Lambda$, are 
due to the
wavefunction renormalization $Z$ of $\Sigma$.\footnote{There 
are corrections
to the K\"ahler potential from higher 
dimensional operators. But, for $v \gg \Lambda$, these are smaller than
the corrections due the wavefunction renormalization \cite{dim1}.}
Thus, the potential for $v$ is:
\begin{equation}
V = \frac{3 \lambda _Q^2 \Lambda^4}{Z(v)}.
\end{equation}
Since $v \gg \Lambda$, we can compute $Z$ in perturbation theory.
The one loop Renormalization Group Equation (RGE) for $Z$ is:
\begin{equation}
\frac{d Z(v)}{d (\ln v)} = \frac{2 Z(v)}{16 \pi^2} \left( 12 g^2_6(v) 
- 6 \lambda _Q ^2 (v)  - \frac{16}{3} \lambda _{\Sigma}^2 (v) \right),
\end{equation}
where $g_6$ is the $SU(6)_{GUT}$ gauge coupling.
The potential can develop a minimum by the inverted hierarchy 
mechanism \cite{witten} as follows. We can choose the 
gauge and Yukawa couplings so that, for large $v$, $\lambda$ dominates
in the above RGE so that $Z(v)$ decreases with
increasing $v$, whereas, for small $v$, $g_6$ dominates so that 
$Z(v)$ increases with $v$. Thus, there is a minimum of $V$ at $v$
such that $\lambda(v) \sim g_6(v)$ so that $dZ(v)/
d(\ln v) =0$.\footnote{This is
a local minimum only
since there is a supersymmetric minimum near the origin 
with $\langle \Sigma \rangle \sim \Lambda \; \hbox{diag}[2,2,-1,-1,-1,-1]$
and $\langle Q \bar{Q} \rangle \sim \Lambda ^2 \;\hbox{diag}
[1,1,-2,-2,-2,-2]$.
However, since 
$v \gg \Lambda$, the tunneling rate from the ``false'' vacuum to this 
global minimum is very small \cite{dim1}.} Due to the 
logarithmic dependence
of $Z,\lambda$ and $g_6$ on $v$, it is possible that at the minimum
$v \gg \Lambda$ which is required for the perturbative 
calculation to be valid.

To get the SM gauge group from the unbroken gauge group,
$SU(3) \times SU(3) \times U(1)$,
we identify one $SU(3)$ with $SU(3)_c$ and we need to break
the (other) $SU(3) \times U(1)$ to $SU(2)_L \times U(1)_Y$.
For achieving this, we use the model in \cite{luty} with a slight 
modification. We next discuss the model.

\section{Specific Models}
Add the following particle content and superpotential:\footnote{Henceforth,
we will suppress the Yukawa couplings in the superpotential.}
\begin{eqnarray}
S & \sim & \; ({\bf 1},{\bf 1}) \nonumber \\
H & \sim & \; ({\bf 6},{\bf 1}) \nonumber \\ 
\bar{H} & \sim & \; ({\bf \bar{6}},{\bf 1})
\end{eqnarray}
\begin{equation}
W_2 = S (H\bar{H} - \Sigma^2). 
\end{equation}
The $F$-flatness condition 
for $S$ forces $H$ and $\bar{H}$ to acquire vev.\footnote{
In \cite{luty}, the terms $S(\hbox{tr} \Sigma ^2 - \Phi ^2)$ and
$T (H \bar{H} - \Phi ^2)$ (where $S,T,\Phi$ are singlets)
were used instead to relate the
$H,\bar{H}$ and $\Sigma$ vev's to the vev of the GUT modulus $\Phi$.} 
We look for a minimum with 
the
vev's of $H,\bar{H}$ in the form:
\begin{equation}
\langle H \rangle = \langle \bar{H} \rangle \sim v \; (1,0,0,0,0,0).
\label{Hvev}
\end{equation}
This breaks $SU(3) \times U(1)$ to $SU(2) \times U(1)$. 

We now discuss the mass spectrum.
The superpotential has a separate $SU(6)$ symmetry acting
on $\Sigma$ and $H,\bar{H}$. The $SU(6)_H$ is broken to $SU(5)$ resulting
in the Nambu-Goldstone fields (with $SU(3)_c \times SU(2)_L$ 
quantum numbers):
\begin{equation}
{\bf (3,1) + (\bar{3},1) + (1,2) + (1,2) + (1,1)}.
\label{6to5}
\end{equation}
The breaking of $SU(6)_{\Sigma}$ to $SU(3) \times SU(3) \times U(1)$ 
generates the Nambu-Goldstone fields:
\begin{equation}
{\bf (3,2) + (\bar{3},2) + (3,1) + (\bar{3},1)}, 
\label{6to331}
\end{equation}
which is the same as Eqn.(\ref{6to33}) but with quantum numbers 
under $SU(3) \times SU(2)$ shown.
The following fields are eaten in the breaking of the $SU(6)$ 
gauge symmetry
to the SM gauge group:
\begin{equation}
{\bf (3,2) + (\bar{3},2) + (3,1) + (\bar{3},1) + (1,2) + (1,2) + (1,1)}.
\label{6toSM}
\end{equation}
The various fields decompose as:
\begin{eqnarray}
\Sigma & \sim & \; {\bf (3,2) + (\bar{3},2) + (3,1) + (\bar{3},1) + 
(8,1) +}  \nonumber \\
 & & {\bf (1,2) +
(1,2) + (1,3) + (1,1) + (1,1)} \nonumber \\
H & \sim & \; {\bf (3,1) + (1,2) + (1,1)} \nonumber \\
\bar{H} & \sim & \; {\bf (\bar{3},1) + (1,2) + (1,1)}.
\end{eqnarray}
As mentioned before, the ${\bf 
(8,1) + (1,2) + (1,2) + (1,3) + (1,1)}$ components of $\Sigma$
(which transform as ${\bf (8,1) + (1,8)}$ 
under $SU(3) \times SU(3)$: see Eqn.(\ref{sigma33})) get a mass
from the $\Sigma^3$ term. The ${\bf (3,2) + (\bar{3},2)}$ 
components of $\Sigma$ and the ${\bf (1,2) + (1,2)}$ 
components of $H,\bar{H}$ are eaten by the broken gauge
symmetry (see Eqn.(\ref{6toSM})). 
From Eqns.(\ref{6to5}) and (\ref{6to331}) there 
are two pairs of Nambu-Goldstone 
triplets in $\Sigma$
and $H,\bar{H}$.
From Eqn.(\ref{6toSM}) only one combination of these  two pairs 
is eaten.\footnote{Without the $H,\bar{H}$, the ${\bf (3,1) +(\bar{3},1)}$
components of $\Sigma$, which along with the 
${\bf (3,2) + (\bar{3},2)}$ components
form ${\bf (3,\bar{3}) + (\bar{3},3)}$ under $SU(3) \times SU(3)$,
are eaten as mentioned before (see Eqns.(\ref{6to33}) and 
(\ref{sigma33})).}
The other combination is massless.
The remaining SM singlet in $\Sigma$ is the flat direction tr $\Sigma^2$.
One combination of the SM singlets in $H,\bar{H}$ is eaten by the broken 
symmetry (see Eqn.(\ref{6toSM})) or in other words is constrained by the
$D$-flatness condition. The other combination is parametrized by
$H\bar{H}$. The singlet $S$ marries one combination of 
tr $\Sigma^2$ and $H \bar{H}$ due to the superpotential $W_2$. The 
orthogonal
combination of $\Sigma^2$ and $H \bar{H}$ is massless. Thus, the massless
fields are this flat direction and a pair of triplets in 
$\Sigma, H$ and $\bar{H}$. 

To make these triplets heavy\footnote{Giving mass to these
Nambu-Goldstone triplets is
equivalent to getting the orientation of the $\Sigma$
and $H,\bar{H}$ vev's in Eqns.(\ref{sigmavev}) and (\ref{Hvev}).}
, we can use the sliding singlet mechanism 
\cite{barr,luty}.
Add the following to the superpotential:
\begin{equation} 
W_3 = H (\Sigma + X ) \bar{h} + \bar{H} (\Sigma + \bar{X} ) h,
\end{equation} 
where 
\begin{eqnarray}
X & \sim & \; ({\bf 1},{\bf 1}) \nonumber \\
\bar{X} & \sim & \; ({\bf 1},{\bf 1}) \nonumber \\ 
h & \sim & \; ({\bf 6},{\bf 1}) \nonumber \\
\bar{h} & \sim & \; ({\bf \bar{6}},{\bf 1}).
\end{eqnarray}
$F_X = F_{\bar{X}} = 0$ forces $h = \bar{h} = 0$. $F_{h} = F_{\bar{h}} = 0$
along with the form of the $H,\bar{H}$ vev's makes the singlets slide
so that $X = \bar{X} = -v/ \sqrt{12}$. Thus, the form of 
the $(\Sigma +X)$ vev is such that
the triplets in $H,\bar{H}$ get a mass with the triplets 
in $h,\bar{h}$.
There is no mass term for the doublets in $h,\bar{h}$ with those in
$H,\bar{H}$.
However, the $H,\bar{H}$ vev's with the above superpotential give
a mass term for the doublets (and also the triplets) 
in $\Sigma$ with those in $h,\bar{h}$ 
(there is also a mass term for the doublets in $\Sigma$ from the 
$\Sigma^3$ term). Also, the $H,\bar{H}$ vev's give mass to 
the first (SM singlet) components of $h,\bar{h}$ with combinations of 
tr $\Sigma^2$ and $X,\bar{X}$. Thus, the only massless field is the flat 
direction which is now a combination of tr $\Sigma^2$, $H\bar{H}, X$ and
$\bar{X}$. Along this flat direction, both SUSY and the GUT 
symmetry are broken.

To get the usual pair of light Higgs doublets, we duplicate the 
above structure of $S, H, \bar{H}, h, \bar{h}, X$ and $\bar{X}$ 
\cite{barr,luty}.
The superpotential is:
\begin{equation}
W_2 + W_3 = \sum_{i=1}^{2} S_i (H_i \bar{H}_i - \Sigma^2) + \sum_{i=1}^{2}
H_i (\Sigma + X_i) \bar{h}_i + \sum_{i=1}^{2} \bar{H}_i 
(\Sigma + \bar{X}_i) {h}_i.
\end{equation}  
$F_{S_2}=0$ forces $H_2 \bar{H}_2 = \Sigma^2$. We look for a minimum with
the vev's of $H_2, \bar{H}_2$ aligned with $H_1,\bar{H}_1$, 
{\it i.e.}, $H_2 =
\bar{H_2} \sim v (1,0,0,0,0,0)$. 
Then, as before, the sliding singlet mechanism gives mass for the triplets
in $H_2,\bar{H}_2$ with those in $h_2,\bar{h}_2$. As before, due to
the vev's of $H_2,\bar{H}_2$, the SM singlets in
$h_2,\bar{h}_2$ get a mass with two combinations of 
tr $\Sigma^2$ and $X_2,\bar{X}_2$. Thus, the flat direction is 
now a combination of
tr $\Sigma ^2$, $H_i \bar{H}_i$, $X_i$ and $\bar{X}_i$ with $i=1,2$.
Only one combination of the 
doublets in $h_1, h_2$ marries the doublet in $\Sigma$ due to the
$\bar{H}$ vev's (similarly for the
doublets in $\bar{h}_{1,2}$). 
This leaves one pair of massless doublets in the $h,\bar{h}$'s
which can be the usual Higgs doublets.
There is also a pair of massless doublets in the $H$'s since only one pair
is eaten in the gauge symmetry breaking (see Eqn.(\ref{6toSM})). 
Also, there is a massless SM singlet
in the $H$'s which can be seen as follows. The $H,\bar{H}$'s have 
four SM singlets.
The $F_S = 0$ conditions relate two combinations of these, namely
$H_1 \bar{H}_1$ and $H_1\bar{H}_2$, to $\Sigma^2$.
One combination is eaten by the broken gauge symmetry (see 
Eqn.(\ref{6toSM}));
in other words, one combination of the vev's is constrained by 
the $D$-flatness
condition. This leaves one combination of the vev's unconstrained, 
{\it i.e.,} one
massless SM singlet in $H,\bar{H}$'s. We discuss two ways to 
give mass to the extra pair of 
doublets and the SM singlet in $H,\bar{H}$.\footnote{
Giving mass to the extra pair of
doublets and the SM singlet in $H,\bar{H}$ is equivalent to getting the 
alignment of the $H_2,\bar{H}_2$ vev's with the $H_1,\bar{H}_1$ vev's.}

In the first model  
we add the 
superpotential $W_4 + W_5$ where:
\begin{equation}
W_4 = \frac{1}{M} \left( \left( H_1 \bar{H}_1 \right) 
\left( H_2 \bar{H}_2 \right) - \left( H_1 \bar{H}_2 \right) 
\left( H_2 \bar{H}_1 \right) \right), 
\label{higherdim}
\end{equation}
with, say, $M=M_{Pl}$
and  
\begin{equation}
W_5 = S_3 \left( H_1 \bar{H}_2 - H_2 \bar{H}_1
\right),
\end{equation}
where $S_3$ is a singlet.
$W_2 + W_3 + W_5$ is invariant under $(H,\bar{H},h,\bar{h},S,X,\bar{X})_1 
\leftrightarrow
(H,\bar{H},h,\bar{h},S,X,\bar{X})_2$
and $S_3 \leftrightarrow
-S_3$ and $W_4$ is invariant under two $SU(2)$'s 
-- one with $\left(H_1,H_2\right)$ as a doublet and the other with
$\left(\bar{H}_1,\bar{H}_2\right)$ as a doublet.
\footnote{Otherwise, we have to tolerate
some fine tuning to get this form of the superpotential.} 
We look for a minimum with $\langle S_3 \rangle = 0$.
$F_{S_3}$ gives an additional 
constraint 
between the $H,\bar{H}$ vev's giving a mass (with $S_3$)
to the SM singlet mentioned above.
The doublets in $H,\bar{H}$ have a mass matrix of the form \cite{barr}:
\begin{equation}
\left( 
\begin{array}{cc}
\langle H_2 \bar{H}_2 \rangle  & -\langle H_1 \bar{H}_2 \rangle \\
- \langle H_2 \bar{H}_1 \rangle  & \langle H_1 \bar{H}_1 \rangle 
\end{array}
\right), 
\end{equation}
which has one zero eigenvalue corresponding to the eaten pair of doublets
and one non-zero eigenvalue $\sim M^2_{GUT}/M$ 
which is the mass for the other pair of
doublets.
This shifts the prediction of $\sin ^2 \theta_W$ by about $+ 
3 \times 10^{-3}$
if $\alpha_s (m_Z)$ and $\alpha_{em} (m_Z)$ are used as inputs.

In the other method \cite{luty},
we add the terms:
\begin{equation}
W_4 ^{\prime} = (X_1 + X_2) 
\Delta ^2 +\left( H_2 \Delta \bar{H}_1 - H_1\Delta  \bar{H}_2 \right), 
\end{equation}
where $\Delta$ is a ${\bf 35}$ of $SU(6)$.
$W_4 ^{\prime}$ is invariant under the symmetry \\ 
$(H,\bar{H},h,\bar{h},S,X,\bar{X})_1 \leftrightarrow
(H,\bar{H},h,\bar{h},S,X,\bar{X})_2$ and $\Delta \leftrightarrow -\Delta$.
We look for a minimum 
with the vev of $\Delta =0$ so that the $F_X$ and 
$F_H$-flatness conditions are not affected. 
The vev's of $X_{1,2}$ give mass to $\Delta$. $F_{\Delta} = 0$ 
gives a constraint between the vev's of the $H,\bar{H}$'s giving a 
mass (with a singlet in
$\Delta$) 
to the SM singlet 
mentioned above. 
Due to the $H,\bar{H}$ vev's, 
the massless pair of doublets in the $H$'s gets a mass
with those in $\Delta$. Thus, the only massless field is the 
flat direction
which breaks both SUSY and the GUT symmetry. 

If we are willing to tolerate fine tuning to ``solve'' the usual 
doublet-triplet splitting
problem to get a pair of light doublets, 
we can gauge only the $SU(5)$ subgroup
of the $SU(6)$. Then, with only the $\Sigma$ field and $W_1$, 
the generators of the global $SU(6)$ in Eqn.(\ref{6to331})
are broken ($SU(6)_{global}$
is broken to $SU(3) \times SU(3) \times U(1)$).
Of these generators, only ${\bf (3,2) + (\bar{3},2)}$ are gauged. 
Thus, $SU(5)_{local}$ is broken down to the SM. We get a pair of massless
triplets in $\Sigma$ corresponding to the broken generators
which are not gauged. These can be given a mass by adding:
\begin{equation}
H ( \lambda_1 \Sigma_1 + \lambda_{24} \Sigma_{24} ) \Sigma_{\bar{5}} + 
\bar{H}
(\bar{\lambda}_1\Sigma_1 + \bar{\lambda}_{24} \Sigma_{24} ) \Sigma_5,
\label{5only}
\end{equation}
where $H,\bar{H}$ are fundamentals of $SU(5)$ and $\Sigma_5, \Sigma_
{\bar{5}}, \Sigma_1
$ and $\Sigma_{24}$ denote components of $\Sigma$ transforming 
as ${\bf 5}, {\bf \bar{5}}, {\bf 1}$ and ${\bf 24}$, respectively,
under $SU(5)$.\footnote{The superpotential
in Eqn.(\ref{w1}) is invariant under the $SU(6)$ global symmetry whereas
the one in Eqn.(\ref{5only}) is only $SU(5)_{local}$ invariant.} Since 
$\langle \Sigma _1
\rangle \sim \hbox{diag}[1,1,1,1,1]$ and $\langle \Sigma_{24} \rangle
\sim \hbox{diag}[-3,-3,2,2,2]$ (in $SU(5)$ space), 
we can fine tune the couplings $\lambda,\bar
{\lambda}$ so that there is a mass term for the triplet in $H \; (\bar{H})$
with the triplet in $\Sigma_{\bar{5}} \; (\Sigma_5)$ but not for 
the doublets.
Then, the doublets in $H,\bar{H}$ can be the usual Higgs doublets.
\footnote{The doublets in $\Sigma_{5,\bar{5}}$ get a mass from the 
$\Sigma^3$ term as before.} 

In all these models, 
the $\mu$ term has to be generated by some mechanism.
Also, these models are only technically natural, {\it i.e.},
the superpotential is not the most general one allowed by symmetries.
For example, in the model with the full $SU(6)$ symmetry gauged, we 
need the
terms $S H \bar{H}$, tr $\Sigma^3$ and $S \; \hbox{tr} \Sigma^2$ and 
so the term
$H \Sigma \bar{H}$ is also allowed which is undesirable. 
So, these models should be viewed as existence proofs
of models in which both a simple GUT gauge group
and SUSY are broken by the same field.

\section{MSSM Spectrum}
\subsection{Quarks and Leptons}
The SM fermion Yukawa couplings can be generated using the
method in \cite{luty} as follows.
Add the following fields charged under $SU(6)_{GUT}$ and superpotential:
\begin{eqnarray}
N_i & \sim & \; {\bf 15} \nonumber \\
\bar{P}_{1i}, \bar{P}_{2i} & \sim & \; {\bf \bar{6}} \nonumber \\
Y & \sim & \; {\bf 15} \nonumber \\
\bar{Y} & \sim & \; {\bf \bar{15}}
\end{eqnarray}
\begin{eqnarray}
W_{Yukawa} & = & N_i (\bar{P}_{1j} \bar{H}_1 + \bar{P}_{2j} \bar{H}_2) + 
N_i (\bar{P}_{1j} \bar{h}_1 + \bar{P}_{2j} \bar{h}_2) \nonumber \\
 & & + N_i N_j Y + (X_1 + X_2) Y \bar{Y} + \bar{Y} (H_1 h_1 - H_2 h_2), 
\end{eqnarray}
where $i,j=1,2,3$ are generation indices. 
This superpotential is invariant under the symmetry 
$(H,\bar{H},h,\bar{h},X)_1 
\leftrightarrow (H,\bar{H},h,\bar{h},X)_2$ and $\bar{P}_1 \rightarrow
i \bar{P}_2$, $\bar{P}_2 \rightarrow i
\bar{P}_1$, $N \rightarrow - i N$, $Y \rightarrow -Y$ and $\bar{Y} 
\rightarrow -\bar{Y}$.
For each generation, the $N \bar{P} \bar{H}$ terms make the ${\bf 5}$ 
(under $SU(5)$) of the $N$
and one combination of the ${\bf \bar{5}}$'s  of $\bar{P}_{1,2}$ heavy, 
leaving the usual ${\bf \bar{5}} + {\bf 10}$ massless.
The $N\bar{P}\bar{h}$ terms give the down quark and lepton Yukawa couplings
whereas the up quark Yukawa couplings arise from the terms
$NNY$ and $\bar{Y} Hh$ after integrating out the $Y,\bar{Y}$ fields. 

\subsection{Sparticle Spectrum}
There is a gauge mediation (GM) contribution to the sparticle masses.
The model has both ``matter'' messengers (the $Q,\bar{Q}$ fields and the
heavy components of $H,h$'s) and ``gauge'' messengers (the heavy
gauge multiplets which have a non-supersymmetric spectrum since the field
breaking the GUT symmetry has a non-zero $F$-component). We compute the 
sparticle spectrum using the method of \cite{giudice}. 
In this method, the scalar (mass)$^2$, $m^2_i$,  are 
computed from the RG scaling of the
wavefunctions of the matter fields and the gaugino masses, $M_A$,
are related to the 
RG scaling of the gauge couplings.
The expressions for the masses are:
\begin{equation}
M_A (\mu) = \frac{\alpha_A (\mu)}{4 \pi} \frac{F_v}{v} \left(b_A - b_6 
\right),
\end{equation}
where $b_A$'s are the beta functions of the SM gauge couplings below 
the GUT
scale and $b_6$ is the beta function of the $SU(6)_{GUT}$ above the
GUT scale, and  
\begin{eqnarray}
m^2_i (\mu)& = & \frac{1}{16 \pi^2} \left( \frac{F_v}{v} \right)^2 
\times \nonumber \\
 & & \left( \sum_A \frac{2 C^i_A}{b_A} \left( 
\alpha_A^2(\mu) \left( b_6 - b_A \right) ^2 - b_6^2 \alpha_6^2 \right) 
+ 2 C^i_6 b_6 \alpha_6^2 \right),
\end{eqnarray}
where $C^i_A$ is the quadratic Casimir invariant
for the scalar $i$ under the gauge
group $A$, {\it i.e.}, $4/3$, $3/4$ for fundamentals of
$SU(3)_c$, $SU(2)_L$  respectively and $3/5 \; 
Y^2$ for 
$U(1)_Y$. $C^i_6 = 35/12$ for fields in ${\bf \bar{5}}$ of
$SU(5)$ (${\bf \bar{6}}$ of
$SU(6)_{GUT}$) and $14/3$ for fields in ${\bf 10}$ of
$SU(5)$ (${\bf 15}$ of
$SU(6)_{GUT}$).   
The beta function for $SU(N_c)$ group is defined as $3 N_c - N_{f,eff}$, 
where  
$N_{eff}$ is the ``effective'' number of flavors. 
$\alpha_6$ is the $SU(6)$
coupling at the GUT scale.
The messengers do {\em not} form complete $SU(5)$ representations and thus
the above mass spectrum is different from the models of gauge mediation
with complete $SU(5)$ multiplets as messengers. For example,
the gaugino masses are {\em not} unified at the GUT scale. 

The above results depend on the beta functions of the SM gauge group
below the GUT scale and the beta function of $SU(6)_{GUT}$ above the
GUT scale.
We assume that
there are no particles with SM quantum numbers between the weak
and the GUT scales so that $b_{1,2,3}$ are
the usual MSSM beta functions. The $SU(6)$ beta function, $b_6$,
depends on the particle content at the GUT scale and thus, in turn, on 
the method used
to generate SM fermion Yukawa couplings and the method used to make
the extra pair of doublets in $H,\bar{H}$
heavy. We consider the case where the above method
is used to generate SM fermion Yukawa couplings and the higher dimensional
operator (Eqn.(\ref{higherdim})) is used to make the extra doublets heavy. 
In this case, the 
beta function $b_6$ (defined as $3 N_c - N_{f,eff}$) is $-11$.
We get $m^2_{\tilde{e} _R} (\mu \sim m_Z) \approx - 8 \times 10^{-4} 
\left( F_v/v \right)^2$,  
whereas all other scalar (mass)$^2$ are positive.
We have to add the supergravity (SUGRA)
contribution to the (mass)$^2 \sim
\left( F_v/M_{Pl} \right)^2$ where $M_{Pl} \sim 2 \times 10^{18}$ 
GeV.\footnote{We
assume that the SUGRA contribution to the (mass)$^2$ is positive.} 
For $v \sim 6 \times 10^{16}$ GeV, 
the two
contributions to $m^2_{\tilde{e} _R}$ are comparable
and thus we can get a phenomenologically
acceptable spectrum.\footnote{
It might seem that this value of $v$ is 
a bit larger than the ``usual'' GUT scale $\sim
2 \times 10^{16}$ GeV. However, as mentioned earlier, the flat direction
$v$ is really a combination of $\sim 7$ fields. If we assume that all
these fields have roughly the same vev, then the vev of each field, in
particular, the $\Sigma,H$ fields is $\sim v/ \sqrt{7}$ which is closer
to the usual GUT scale.}
However, since the supergravity contribution
is comparable to the flavor blind GM contribution, we need to impose
some flavor symmetries or alignment 
(of the SUGRA contribution with the
Yukawa couplings) to avoid too large SUSY contributions to FCNC's.
For the squarks, the GM contribution is larger so that less degeneracy
is required in the SUGRA contribution.

\section{Inverted Hierarchy}
Since the flat direction is a combination of the fields tr $\Sigma^2,
H_i\bar{H}_i, X_i$ and $\bar{X}_i$ ($i=1,2$), the RGE analysis
for the wavefunction of the flat direction involves too many
Yukawa couplings. To simplify the analysis, we assume the
the vev's of all the fields in the flat direction are of the same order and
that among the Yukawa couplings, only the $\Sigma Q \bar{Q}$
coupling is large. The $SU(6)_{GUT}$ coupling at the Planck scale is fixed
with the assumption of a desert between the weak and the GUT scales and the
particle content at the GUT scale. We require $\left( F_v /v \right) \sim
10$ TeV to get the sparticle masses $\sim 100$ GeV to $1$ TeV. With
$v \sim 10^{16}$ GeV, this determines $F_v \sim \Lambda^2$ and hence
the $SU(6)_S$ gauge coupling at the Planck scale.
Then, we checked that for the $\Sigma Q \bar{Q}$ coupling
$\sim 2$ at the Planck scale, 
we do get a minimum of the potential at around the GUT scale.

There is also a SUGRA contribution to the (mass)$^2$ $\sim 
\left( F_v/M_{Pl} 
\right)^2$ of the flat direction. 
For $v \sim 10^{16}$ GeV, we expect this to be comparable to the 
(mass)$^2$ due to the inverted hierarchy which is
$\sim -F_v^2 /v^2 \; d^2 Z(v)/ d (\ln v)^2$.
It turns out that in this case the SUGRA contribution is 
smaller (by a factor of $\sim 4$)
than the (mass)$^2$ due to the inverted hierarchy.
This results in a shift of the minimum of $v$ by $\sim O(1/4) \;v$.

To summarize, we have presented a model in which the field breaking
SUSY is the same as the field which breaks a simple GUT gauge group 
to the SM
gauge group. The model is calculable -- it uses the inverted hierarchy
mechanism to generate a minimum for the field in a perturbative
region.
As far as we know, this is the first example of such a kind in the 
literature.

\section{Acknowledgements}
The author would like to thank Nima Arkani-Hamed,
Csaba Cs\'aki, Michael Graesser,
Stephen Hsu, 
Takeo Moroi, 
Hitoshi Murayama 
and John Terning 
for useful discussions.
This work was supported in part by the Director, Office
of Energy Research,
Office of High Energy and Nuclear
Physics, Division of High Energy Physics of the
U.S. Department of Energy under Contract DE--AC03--76SF00098 and in part
by DOE grant number DE-FG03-96ER40969.


\begin{thebibliography}{99}
\bibitem{cheng} H.-C. Cheng, 
Phys. Lett. {\bf B410}, 45 (1997).
\bibitem{graesser} M. Graesser, hep-ph/9805417.
\bibitem{goldberg} H. Goldberg, Phys. Lett. {\bf B400}, 301 (1997).
\bibitem{kolda} C. Kolda, N. Polonsky, 
Phys. Lett. {\bf B433}, 323 (1998).
\bibitem{luty} Z. Chacko, M. A. Luty, E. Ponton, hep-ph/9806398.
\bibitem{jap} T. Hirayama, N. Ishimura, N. Maekawa, hep-ph/9805457.
\bibitem{dim1} S. Dimopoulos, G. Dvali, R. Rattazzi, G. Giudice,
Nucl. Phys. {\bf B510}, 12 (1998). 
\bibitem{witten} E. Witten, Phys. Lett. {\bf B105}, 267 (1981).
\bibitem{barr} S. M. Barr,
Phys. Rev. {\bf D57}, 190 (1998).
\bibitem{giudice} G. Giudice, R. Rattazzi,
Nucl. Phys. {\bf B511}, 25 (1998). 
\end{thebibliography}
\end{document}